\documentclass[twocolumn, a4paper, pra]{revtex4}

\usepackage{graphicx, epsfig, fancybox}
\usepackage{amssymb,amsmath}
\usepackage{color}

\usepackage{times}

\def \del{\partial}    % for writing partial derivatives
\def \hf{\tfrac{1}{2}}    

\def \ord{\mathcal{O}}

\def\lba{\left(}    \def\rba{\right)}
\def\lbc{\left[}    \def\rbc{\right]}

\newcommand{\bra}[1]{\langle\left.{#1}\right|}
\newcommand{\ket}[1]{\left|{#1}\right.\rangle}
\newcommand{\xpct}[1]{\langle{#1}\rangle}    % expectatn value

\newcommand{\ttau}{\tilde{\tau}} 

\begin{document}

\title{Slow interaction ramps in trapped many-particle systems: universal
  deviations from adiabaticity}
% \title{Non-adiabatic interaction ramps in trapped Bose condensate}

\author{Masudul Haque}
\author{F.~E.~Zimmer}

\affiliation{Max Planck Institute for the Physics of Complex Systems, N\"othnitzer Str.~38, 01187 Dresden, Germany}

\begin{abstract}

For harmonic-trapped atomic systems, we report system-independent non-adiabatic features in the
response to interaction ramps.  We provide results for several different systems in one, two, and
three dimensions: bosonic and fermionic Hubbard models realized through optical lattices, a
Bose-Einstein condensate, a fermionic superfluid and a fermi liquid.  The deviation from
adiabaticity is characterized through the heating or excitation energy produced during the ramp.  We
find that the dependence of the heat on the ramp time is sensitive to the ramp protocol but has
aspects common to all systems considered.  We explain these common features in terms of universal
dynamics of the system size or cloud radius.

\end{abstract}

%\pacs{???, ???, ???}


%% 67.85.-d   Ultracold gases, trapped gases
%% 67.85.De   Dynamic properties of condensates; excitations, and superfluid flow 
%% 67.85.Jk   Other Bose-Einstein condensation phenomena
%%
%% 03.75.-b   Matter waves 
%% 03.75.Kk   Dynamic properties of condensates; collective and hydrodynamic excitations, superfluid flow
%% 03.75.Nt   Other Bose-Einstein condensation phenomena
%%
%% 67.10.-j   Quantum fluids: general properties
%% 67.10.Fj   Quantum statistical theory
%%
%% 05.70.-a   Thermodynamics 
%% 05.70.Ln   Nonequilibrium and irreversible thermodynamics
%%
%% 75.10.Pq   Spin chain models
%% 03.65.Xp   Tunneling, traversal time, quantum Zeno dynamics
%% 03.67.Hk   Quantum communication
%% 75.10.Jm   Quantized spin models
%% 75.45.+j   Macroscopic quantum phenomena in magnetic systems

\maketitle

\section{Introduction} 

Adiabaticity is an essential and ubiquitous concept in quantum dynamics.
In the current era of 
% With the advent of
many novel non-equilibrium experimental possibilities,
% through cold-atom or nanoscience developments, 
\emph{deviations} from adiabaticity in slow parameter changes have attracted a lot of attention
\cite{finiteRateQuenches_reviews, Polovnikov_AdiabaticPertThy,finiteRateQuenches_residualenergy, EcksteinKollar_NJP2010, CanoviRossiniFazioSantoro_JSM09,MoeckelKehrein_NJP2010, DoraHaqueZarand_arxiv10, DziarmagaTylutki_arxiv11, Venumadhav_PRB2010,ramp_papers_finitesystems}.
The question of non-adiabaticity is of fundamental interest, but also has practical implications.
Many experimental protocols involve adiabatically changing a parameter in order to reach a desired
quantum state.  Since non-adiabatic heating can rarely be completely avoided, it is essential to
understand deviations from adiabaticity in slow ramps.
%
%% In addition, the proposal of \emph{adiabatic quantum computation}
%% \cite{AdiabaticQC} raises the question of how a realistic parameter ramp in a
%% quantum system deviates from adiabaticity.
% 
While the effect of quantum critical points in the ramp path has been considered in much
detail \cite{finiteRateQuenches_reviews,Polovnikov_AdiabaticPertThy}, settings for non-equilibrium
experiments in isolated systems tend to be mesoscopic rather than macroscopic, without true quantum
critical points.
%
%% first prototypes of such a quantum computer will presumably be mesoscopic
%% rather than macroscopic, without true quantum critical points.
% 
Understanding non-adiabatic ramps in \emph{finite} quantum systems is therefore vital.
Also, since cold atoms dominate experimental non-equilibrium studies, a
harmonically trapped many-particle system is the most important paradigm today
for studying quenches and ramps.
A few studies of ramps in finite and trapped systems have appeared in the very
recent literature \cite{Venumadhav_PRB2010, ramp_papers_finitesystems},
indicating an emerging recognition of the importance of the adiabaticity issue
in finite systems.
In addition, in the past couple of years, reports of experimental
investigations of finite-rate interaction ramps have started to appear, both
in the continuum \cite{Salomon_BEC_ramp_2011} and in optical-lattices
\cite{HungZhangGemelkeChin_PRL10, BakrPolletGreiner_Science2010,
  DeMarco_PRL11}, with harmonically trapped atoms.

In this work, we consider non-adiabatic ramps in several
\emph{distinct} interacting many-particle systems, confined in isotropic
harmonic traps.  In each case, we consider ramps of the interaction from an
initial value $U_{\rm i}$ to a final value $U_{\rm f}$, occurring in time scale $\tau$.
We focus on large but finite $\tau$ (near-adiabatic ramps).  We study
deviations from adiabaticity through the heating $Q$, which is the final
energy at time $t\gg\tau$ minus the ground state energy of the final
Hamiltonian.  This quantity is also called the residual energy or excess
excitation energy \cite{EcksteinKollar_NJP2010,
CanoviRossiniFazioSantoro_JSM09, finiteRateQuenches_residualenergy,
Venumadhav_PRB2010}, and may be thought of as the ``friction'' due to
imperfect adiabaticity \cite{Muga_delCampo_frictionless}.
The asymptotic form of $Q(\tau)$ is a quantitative characterization of minimal
corrections to adiabaticity.
Ref.~\cite{DeMarco_PRL11} reports measurements of excess energies after ramps,
which may be regarded as the first experimental approach to this quantity.

We find that the asymptotics of $Q(\tau)$ is common to all the compressible
systems that we considered.
%
%% We find that the asymptotics of $Q(\tau)$ is common to all the compressible
%% systems that we considered: we have thus uncovered \emph{universal}
%% non-adiabatic features of ramp dynamics in trapped systems.  
%
The $Q(\tau)$
function has overall power-law decay, $Q\sim\tau^{-\nu}$, with the exponent
$\nu$ depending on the shape of the ramp.  For certain ramp shapes, $Q(\tau)$
has oscillations superposed on top of the power-law decay.
%
%% The oscillation frequency is given by the breathing-mode frequency determined
%% mostly by the harmonic trap.  (The dependence on the interaction strength is
%% often weak.)
%
We present results for a range of systems, interactions, and dimensionalities,
which make clear that the universal features are independent of system details
and generally do not depend on the initial and final values of the
interaction, as long as the trapped system remains in the same phase.

Since the effects are universal over a wide range of trapped systems, they
should be due to some type of dynamics that is prevalent in many harmonically
trapped systems.  We show that the relevant dynamics is the size oscillation
or breathing-mode oscillation of the trapped cloud.  Almost all trapped
systems have ``soft'' breathing modes due to vanishing density at the edge,
irrespective of the nature of intrinsic modes of the system.  Our results show
that, in the slow ramp limit, the breathing modes due to trapping dominate the
near-adiabatic response of many systems.  To show that the asymptotic
behaviors of $Q(\tau)$ are due to size dynamics, we will use a variational
description for one of the systems (the Bose condensate), treating the extent
(radius) of the many-particle cloud as a time-dependent variational parameter.
Such a ``radius dynamics'' description will be shown to reproduce the
universal $Q(\tau)$ behaviors in thorough detail.
An equivalent formulation is not easy to set up for all the systems;
nevertheless, the commonality of the $Q(\tau)$ features, and the success of
the radius description for at least one case, is convincing argument that the
same dynamics type is the relevant feature in each case.

Trapped atoms are by far the most promising setup for experimentally exploring isolated-system
dynamics in general and non-adiabaticiy issues in particular.  A universal excitation mechanism that
is dominant in generic trapped systems is thus an important baseline perspective for understanding
the many further non-adiabatic ramp experiments expected in the near future.
Contemporary theoretical treatments of non-adiabaticity in many-particle systems almost invariably
appeal to cold-atom experiments for motivation.  Yet, our results show that in a real trapped-atom
experiment, radius dynamics dominates over the intrinsic heating mechanisms that may be important in
individual uniform systems.  In other words, the power-laws and many other results that have become
available in the literature for $Q(\tau)$ for uniform systems (e.g.,
Refs.\ \cite{finiteRateQuenches_reviews, Polovnikov_AdiabaticPertThy,
  finiteRateQuenches_residualenergy, EcksteinKollar_NJP2010, CanoviRossiniFazioSantoro_JSM09,
  MoeckelKehrein_NJP2010, DoraHaqueZarand_arxiv10, DziarmagaTylutki_arxiv11}) will be either absent
or hidden in any trapped-atom experiment designed to see such behaviors.

After introducing the different systems concisely in Sec.\ \ref{sec_systems} and the shapes of the
interaction ramps in Sec.\ \ref{sec_rampshapes}, we give a description of the generic $Q(\tau)$
features in Sec.\ \ref{sec_universal_features}.  Sec.\ \ref{sec_radius_dynamics_analysis} provides a
analysis of the size (radius) dynamics and shows how size dynamics explains the common $Q(\tau)$
features.  Sec.\ \ref{sec_spectral} uses the perturbative results of
Ref.\ \cite{EcksteinKollar_NJP2010} and arguments about the spectrum of trapped many-body systems to
derive the same $Q(\tau)$ features in the perturbative (small-quench) regime.  We provide context
and point out some open questions in Sec.\ \ref{sec_discussion}.  The Appendix gives further details
about the different methods used to treat the different systems.

\section{Distinct trapped systems \label{sec_systems}}

We will present results for fermionic and bosonic systems, with and without optical lattices.  Here
we present concisely the systems and the methods used for calculating time evolution.  Additional
detail is provided in the Appendix.

\paragraph*{Lattice systems.}

The lattice systems are two-component fermions, and single-component bosons, described respectively
by the fermionic and bosonic Hubbard models \cite{Lewenstein_review}:
\begin{gather*}
H_{fH} = - \sum_{<ij>\sigma} \lba c_{i\sigma}^{\dagger}c_{j\sigma}+h.c.\rba  + U\sum_{j}
n_{j\uparrow}n_{j\downarrow} + H^{\rm tr} \ ;  \\
H_{bH} =  - \sum_{<ij>} \lba b_i^{\dagger}b_j + h.c. \rba  + U\sum_{j}
n_j\lba{n_j-1}\rba + H^{\rm tr} \ .
\end{gather*}
We use the (inverse) hopping strength as energy (time) units, which is why the
hopping terms (first terms in each Hamiltonian) have unit coupling
coefficient.  As usual, $c_{j\sigma}^{\dagger}$ ($b^{\dagger}$) are fermionic
(bosonic) creation operators, $\sigma$ is a spin index, and $\xpct{ij}$
indicates nearest-neighbor sites.
%
%% In one dimension (1D), the trap terms are $H^{\rm tr}=\hf k_{\rm tr} \sum
%% {j^2}c_{j\sigma}^{\dagger}c_{j\sigma}$ for fermions and $\hf k_{\rm tr} \sum
%% {j^2}b_{j}^{\dagger}b_{j}$ for bosons.
%
The trap terms are 
\[
H^{\rm tr} = \hf k_{\rm tr} \sum_{j\sigma} {j^2}c_{j\sigma}^{\dagger}c_{j\sigma} \quad \mathrm{and} \quad
H^{\rm tr} = \hf k_{\rm tr} \sum_{j} {j^2}b_{j}^{\dagger}b_{j}
\]
for fermions and for bosons respectively, for one dimension (1D).
We will consider both spin-balanced ($N_{\uparrow}=N_{\downarrow}$) and
polarized ($N_{\uparrow}{\neq}N_{\downarrow}$) fermionic systems.  
We treat time evolution of the lattice systems through exact numerical
evolution of the full quantum wavefunction, for few-particle configurations in
1D.  The Bose-Hubbard model is also treated via the Gutzwiller approximation \cite{Lewenstein_review},
for $N>10$ bosons and/or for 2D.

%% [

%% OR: For one dimension ($D=1$), we treat few-particle instances of both systems
%% through numerically exact evolution of the full quantum wavefunction.  The
%% Bose-Hubbrard model (all $D$) is also treated using the time-dependent
%% Gutzwiller formulation (cite), which is not exact but allows treatment of
%% larger systems. 

%% ]

\paragraph*{Continuum systems}

The continuum systems (without optical lattice) are the Bose-Einstein
condensate (BEC) and the interacting two-component Fermi gas.  We use trap
units for these cases, expressing lengths (energies) in units of trap
oscillator length (trapping frequency).

\paragraph*{Continuum: (1) Bose condensate.}

The BEC is treated via the Gross-Pitaevskii (GP) description \cite{pitaevskii-jetp13,gross-nc20}.
The dynamics is given by the time-dependent GP equation,
% \cite{pitaevskii-jetp13,gross-nc20},
$i\frac{\partial\psi}{\partial t} =
-\hf\bigtriangledown^2\psi+\tfrac{1}{2}r^2 \psi+U(t)|\psi|^2\psi$, and the GP
energy functional is 
\begin{equation*}
E[\psi] ~=~ \int_{r} \, \lbc -\hf\psi^* \nabla^2\psi  ~+~ \hf U(t) |\psi|^4
~+~ \hf r^2 |\psi|^2 \rbc \, .
\end{equation*}
Here $r$ is the radial position; $\int_r\equiv{\int}d^Dr$ is the spatial
integral for dimensionality $D$; $U$ is the effective interaction strength
whose relation to the physical interaction is also $D$-dependent (c.f.\
Ref.~\cite{Olshanii_PRL98} for 1D).
%
%% The GP equation provides an excellent account of many aspects of trapped
%% condensate dynamics.  Although physics beyond GP is more important in lower
%% dimensions, it is natural to express GP results general to all $D$, which we
%% do.
%
We normalize $\int_r|\psi|^2=1$, so that $U$ contains a factor of the boson
number $N$; thus $U$ can be large even within the mean-field regime.  
We will present GP results general to all $D$.

In addition to full solutions of the GP equation, we also use a variational
description which uses the radius or size of the condensate, $\sigma(t)$, as the
time-dependent parameter.
Using a gaussian ansatz for the cloud shape, the equation of motion for
$\sigma$ is found to be
\begin{equation}
\sigma{\sigma''}  + \sigma^2 - \sigma^{-2}  -  (\sqrt{2\pi}\sigma)^{-D} U(t)
~=~ 0   \, , 
\label{eq:variational_radiusEq} 
\end{equation}
%
%% \begin{equation}
%% \sigma\frac{{\rm d}^2\sigma}{{\rm
%%     d}t^2}+\sigma^2-\frac{1}{\sigma^2}-\frac{U(t)}{(\sqrt{2\pi}\sigma)^D}=0 \, , 
%% \label{eq:variational_radiusEq} 
%% \end{equation}
%
the primes denoting time derivatives.  The energy is
\begin{equation}
E[\sigma] =   \tfrac{1}{4}D\left[
\sigma^{-2}+\sigma^2 + (\sigma')^2 \right]
~+~ \hf (\sqrt{2\pi}\sigma)^{-D} U(t) \, .
\label{eq:variational_energyEq}
\end{equation}
%
%% \begin{equation}
%% E[\sigma] =   \frac{D}{4}\left[
%% \frac{1}{\sigma^2}+\sigma^2 + \left(\frac{{\rm d}\sigma}{{\rm d}t}\right)^2 \right]
%% ~+~ \frac{U(t)}{2(\sqrt{2\pi}\sigma)^D} \, .
%% \label{eq:variational_energyEq}
%% \end{equation}
%
This radius description is suitable for describing breathing-mode
oscillations.  For constant $U$, for small-amplitude oscillatory solutions
$\sigma = R_0+\rho\sin(\Omega_B{t})$ (amplitude $\rho$),
Eq.\ \eqref{eq:variational_energyEq} shows that the excitation energy scales
as ${\sim}\rho^2$.
%
% ; we will use this for our energy analysis below.
%
We could also use a Thomas-Fermi instead of Gaussian profile; the results are
very similar and do not affect our arguments.

\paragraph*{Continuum: (2) weakly interacting 2-component fermions.}

The continuum fermionic system (3D) is treated using a hydrodynamic description.  There are several
similar formulations; we choose the so-called ``time-dependent DFT'' \cite{fermions_tddft}, where
the fermionic gas is described by a nonlinear Schr\"odinger equation, $i\frac{\partial\psi}{\partial
  t} = [-\hf\nabla^2 +\tfrac{1}{2}r^2 +\mu(n,U)]\psi$.  Here $n=|\psi|^2$ is the sum of the (equal)
densities of the two components.  We use the Hartree expression:
%
% $\mu(n,U) = \hf(3\pi^2n)^{2/3} + {\hf}Un$. 
%
\begin{equation}  \label{eq:Hartree_chempot}
\mu(n,U) = \hf(3\pi^2n)^{2/3} + {\hf}Un  \, . 
\end{equation}
Both a paired superfluid ($U<0$) and a Fermi liquid ($U>0$) are described by
the same formalism.  
Analogous to the BEC case, we use a variational description to formulate
the dynamics in terms of the cloud radius.
%
% ; we will present results using this method.

\begin{figure}
\centering
\includegraphics*[width=0.99\columnwidth]{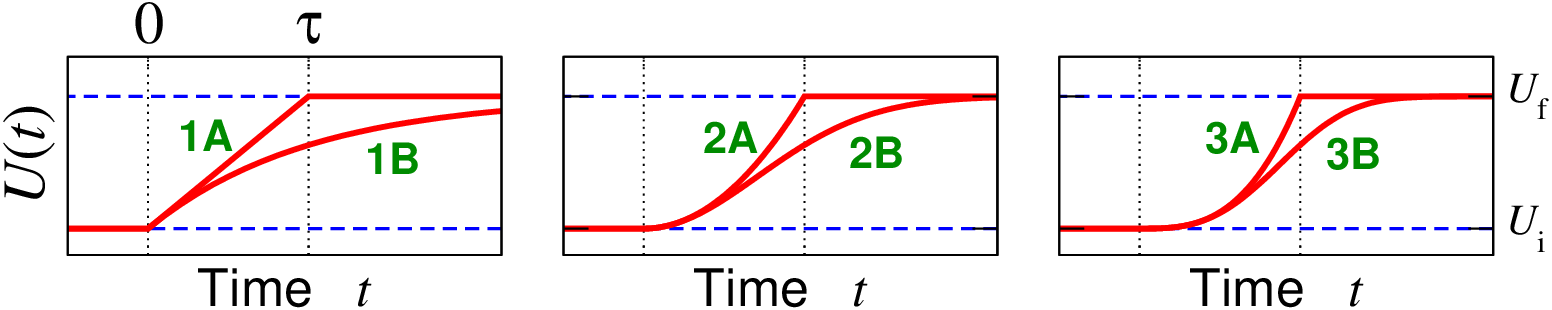}
\caption{ \label{fig_rampshapes}
%
% Ramp shapes [1A], [1B], \& [2A], [2B].   
%
Ramp shapes. 
We illustrate $U_{\rm f}>U_{\rm i}$; it should be straightforward to visualize
$U_{\rm f}<U_{\rm i}$ cases.
Each [A], [B] pair has the same initial behavior --- linear, quadratic and
cubic for the three pairs.  The [B] versions have no endpoint kinks. 
}
\end{figure}

\section{Ramp shapes \label{sec_rampshapes}} 

We analyze interaction ramps of the form
\[
U(t) ~=~  U_{\rm i} ~+~  \theta(t)\, (U_{\rm f}-U_{\rm i})\, s(t/\tau)  \, .
\]
The ramp function $s(x)$ starts at $s(0)=0$ and ends at $s(\infty)=1$.
The ramps take place over time scale $\tau$ but we do not require them to end at $t=\tau$.
(Contrast, e.g., Ref.~\cite{EcksteinKollar_NJP2010}.)
We choose a collection of ramps which allows us to compare the presence/absence of kinks and various
exponents.

Specifically, we consider the following forms for $s(x)$:
%
%% \begin{align*}
%% {\rm [1A]} \quad  s(x) &= x\; \theta(1-x)  ~+~ \theta(x-1) &  {\rm [1B]} \quad  s(x)&= 1-e^{-x} \\
%% {\rm [2A]} \quad  s(x) &= x^2\; \theta(1-x)  ~+~ \theta(x-1) &  {\rm [2B]} \quad  s(x)&= 1-e^{-x^2} \\
%% {\rm [3A]} \quad  s(x) &= x^3\; \theta(1-x)  ~+~ \theta(x-1) &  {\rm [3B]} \quad  s(x)&= 1-e^{-x^3} 
%% \end{align*}
%
\begin{align*}
&{\rm [1A]} \quad  x\; \theta(1-x)  +\theta(x-1)  & {\rm [1B]} \quad  & 1-e^{-x} \\
&{\rm [2A]} \quad  x^2\; \theta(1-x)  +\theta(x-1)  & {\rm [2B]} \quad & 1-e^{-x^2} \\
&{\rm [3A]} \quad  x^3\; \theta(1-x) + \theta(x-1) & {\rm [3B]} \quad  & 1-e^{-x^3} 
\end{align*}
Each [A], [B] pair has the same initial behavior, $s(x){\sim}x^{\alpha}$, but
the [B] versions have no endpoint kinks (Fig.\ \ref{fig_rampshapes}).

%% Our main object of study is the residual energy as a function of ramp time,
%% i.e., the $Q(\tau)$ function.

\begin{figure}
\centering
\includegraphics*[width=0.99\columnwidth]{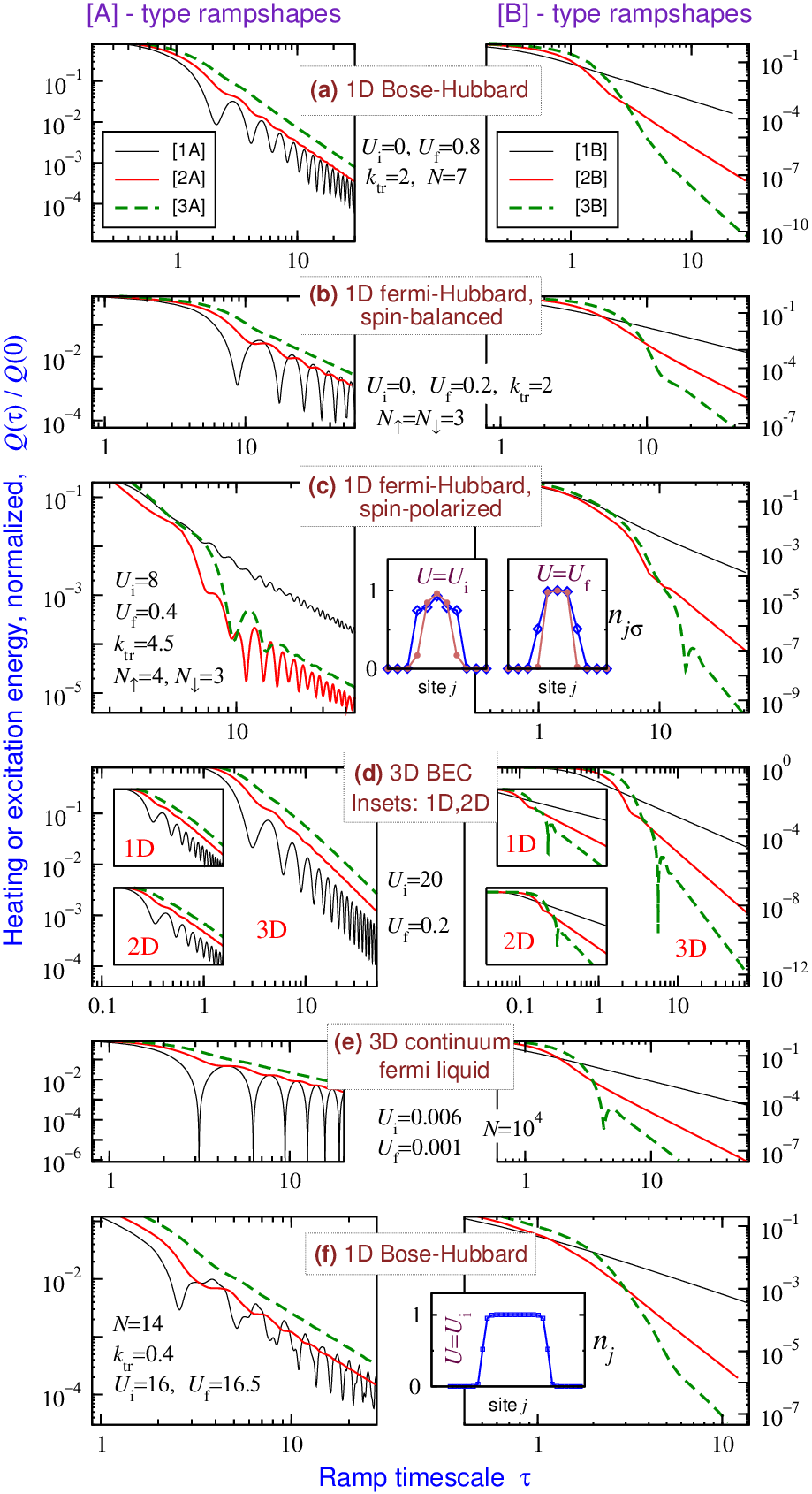}
\caption{ \label{fig_heatcurves}
%
% (Color online.)
%
The normalized excitation energy or heat, $Q(\tau)/Q(0)$, for various trapped systems.
($U_{\rm i}$,$U_{\rm f}$) values and trap constants $k_{\rm tr}$ are chosen over a wide
range to emphasize the parameter-independence of the effects.
Response to [A] ramps (left panels) shows overall $Q(\tau)\sim\tau^{-2}$
decay, with superposed oscillations that decay faster for larger-$\alpha$
ramps.  For [B] ramps (right panels) the asymptotic behavior is pure
$\tau^{-2\alpha}$ decay.
Insets to (c): density profiles for $U=U_{\rm i,f}$ ground states.  For (f),
 $U=U_{\rm i,f}$ ground-state profiles are very similar, only $U_{\rm i}$
 shown.
}
\end{figure}

\section{Universal features of ramp response \label{sec_universal_features}}

In Fig.\ \ref{fig_heatcurves} we present the behavior of the heat
function $Q(\tau)$, normalized against its instantaneous-quench value
$Q(\tau=0)$.
The small-$\tau$ behavior and the exact magnitude of $Q(\tau)$ are system- and
approximation-dependent; the universal features we present concern only the
large-$\tau$ asymptotics of  $Q(\tau)/Q(0)$.  

In the left panels, the $Q(\tau)$ behaviors are compared for the ramp shapes
$s(x){\sim}x^{\alpha}$ with discontinuous derivatives at endpoints, [1A], [2A],
[3A].  Each curve has an overall power-law decay
\emph{with the same decay exponent}, $Q(\tau)\sim\tau^{-2}$.  This suggests
that the residual energy for such ramps is primarily set by the endpoint kink.
%
% ; we will later show this to be the case using radius-oscillation considerations.  
%
Superposed on the power-law decay are oscillations, which are often but not
always smaller for larger $\alpha$ (contrast panels (c) with others).
The oscillation strength decays faster for larger $\alpha$, as $\sim\tau^{-(\alpha+1)}$.

The right panels concern smoothed ramps [1B], [2B], [3B], which lead to
non-oscillating decay of $Q(\tau)$.  The decay exponent is seen to depend on
the power $\alpha$ of $s(x)\sim{x^\alpha}$, namely,
$Q(\tau)\sim\tau^{-2\alpha}$.

The dimensionality does not affect the decay exponents or general behavior.
As an example, results are shown for the BEC in 1D, 2D, and 3D, in the (d)
panels and insets.
We have also found that the same exponents and oscillation features also
appear in additional cases not shown, e.g., a continuum Fermi superfluid (3D,
$U<0$), the Bose-Hubbard model in higher $D$ (treated via the Gutzwiller
approximation), etc.

The (c) and (f) panels involve systems which cannot be described as having
single-radius profiles.  The two spin components have different extents in the
spin-imbalanced Hubbard model (c); the Bose-Hubbard situation (f) has
superfluid wings around a Mott core.  The $Q(\tau)$ behaviors are more rich
for these systems; however, the features discussed above remarkably also
persist in these more complex cases.

\section{Radius dynamics interpretation \label{sec_radius_dynamics_analysis}}

\begin{figure}
\centering
\includegraphics*[width=0.99\columnwidth]{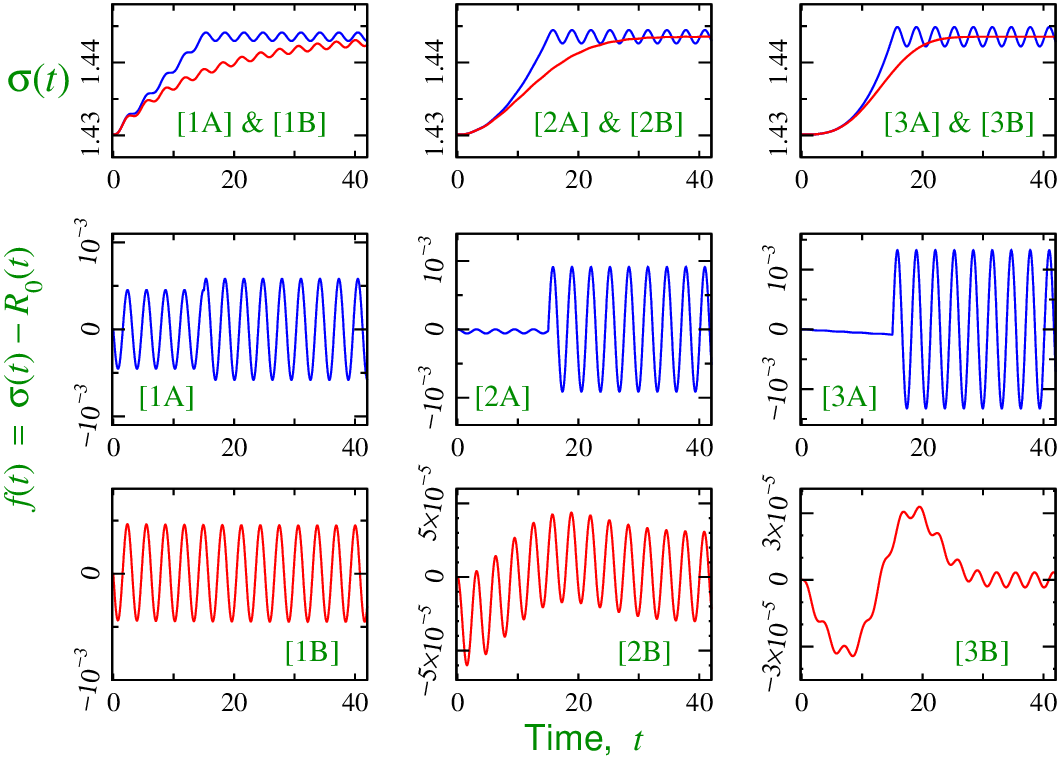}
\caption{ \label{fig_radusOscillatns}
Top row: radius dynamics $\sigma(t)$ for various ramp shapes,
$\tau=15$.  Center and bottom rows: deviation $f(t)$ from the `instantaneous'
ground-state radius $R_0(t)$.
2D BEC;  ($U_{\rm i}$,$U_{\rm f}$) = (20,21).
}
\end{figure}

We now show how radius dynamics explains the $Q(\tau)$ behaviors presented
above.  We will use the convenient formulation of BEC radius dynamics,
Eqs.\ \eqref{eq:variational_radiusEq} and \eqref{eq:variational_energyEq}. 
Fig.\ \ref{fig_radusOscillatns} (top row) shows the radius of a 2D BEC
evolving as a function of time for various ramp shapes, for reasonably large
$\tau$.  In the center and bottom rows, we show the deviation of $\sigma(t)$
from the equilibrium radius corresponding to the instantaneous value of the
interaction, $R_0(t) = R_0[U(t)]$.  For a truly adiabatic ramp, $\sigma(t)$
would follow $R_0(t)$ exactly; therefore the deviation $f(t)=\sigma(t)-R_0(t)$
is at the heart of non-adiabaticity and this quantity determines the heating
$Q$.  After the ramp, the $f(t)$ function is purely oscillatory; $Q$ scales as
the square of the oscillation magnitude.
%
%% Fig.\ \ref{fig_radusOscillatns} presents radius dynamics for $D=2$; the 1D
%% and 3D cases are very similar.
%
The oscillations initiated at the beginning of the ramp are of magnitude
$\ord(\tau^{-\alpha})$ for the $(t/\tau)^{\alpha}$ ramp.  In the case of [A]
ramps (middle row), the endpoint kink causes an $\ord(\tau^{-1})$ oscillation,
which is parametrically larger for $\alpha>1$ and hence dominates the final
dynamics, leading to overall $Q(\tau)\sim\tau^{-2}$ behavior.

%% For the [A] ramps (middle row), the final oscillation strength of $f(t)$ is
%% determined at the endpoint kink.  For the $(t/\tau)^{\alpha}$ ramp, the
%% oscillation magnitude is $\ord(\tau^{-\alpha})$ during the ramp, and turns
%% into $\ord(\tau^{-1})$ after the endpoint kink.  For $\alpha>1$, the final
%% $\ord(\tau^{-1})$ oscillation is parametrically larger than the during-ramp
%% $\ord(\tau^{-\alpha})$ oscillation.

\paragraph*{[A] ramp shapes.}

We first explain the $\sim\tau^{-1}$ scaling of oscillations initiated at the
kink.
If we neglect the smaller oscillations at $t<\tau$, the radius
$\sigma(t){\approx}R_0(t)$ at the kink $t=\tau$ has ``correct'' value for
$U=U_{\rm f}$, i.e. $f$ is negligible.
% and $\sigma(\tau){\approx}R_0[U_{\rm f}]$.   
%
However the derivative is nonzero, $\sigma'(t) {\approx} R_0'(t)|_{t=\tau}$,
which scales as $\sim\tau^{-1}$.  
Thus we have the following ``initial'' conditions at $t=\tau^+$ for subsequent
evolution: $f(\tau)=0$, $f'(\tau^+)=c_0\tau^{-1}$.
Using $f(t>\tau)\approx\rho\sin(\Omega_B{t}+\delta)$, these initial values imply
$\rho\sim\tau^{-1}$.  This explains the $\ord(\tau^{-1})$ oscillation
magnitude of $f(t>\tau)$, and hence $\ord(\tau^{-2})$ heating, for ramps
having an endpoint kink.

The oscillations of $Q(\tau)$ (Fig.\ \ref{fig_heatcurves} left panels) can be
explained by relaxing the approximation $f(t<\tau)\approx0$ made above.
%
% In fact $f(t<\tau)$ is not zero, only parametrically small for large $\tau$.
%
The small oscillations of $f(t<\tau)$ guarantee that $\sigma'(t=\tau)$ oscillates around
$R_0'(t=\tau)$ as a function of $\tau$.  This causes the final breathing mode amplitude $\rho$ to
oscillate around its $\ord(\tau^{-1})$ value as a function of $\tau$.  The $f(t<\tau)$ oscillation
strength is $\ord(\tau^{-\alpha}$) (shown below).  As a result, if
$f(t>\tau)\approx\rho\sin(\Omega_B{t}+\delta)$, we will have
\[
\rho ~\sim~ \frac{c_1}{\tau} + \frac{c_2T(\tau)}{\tau^{\alpha}}
\]  
where $T(x)$ is an oscillatory function.  Therefore, the excess energy ($\sim\rho^2$) has an
$\ord(\tau^{-\alpha-1})$ oscillatory correction to the leading $\ord(\tau^{-2})$ decay.  The
oscillations in $Q(\tau)$ therefore decay as $\tau^{-(\alpha+1)}$, as seen in
Fig.\ \ref{fig_heatcurves} left panels.

\paragraph*{Smoothed [B] ramp shapes.}

For smooth [B] ramps (Fig.\ \ref{fig_radusOscillatns} bottom), the
breathing-mode strength ($\sim\tau^{-\alpha}$) initiated at the beginning of
the ramp remains unchanged; there is no kink to abruptly create larger
oscillations.
We therefore need only to explain the strength of oscillations at the
beginning of the ramp, where $s(t/\tau) \approx (t/\tau)^{\alpha}$.
%
%% (The envelope of the oscillations are somewhat dramatic for larger $\alpha$
%% [3A], but flattens to zero as the interaction levels off to $U_{\rm f}$ at
%% $t\gg\tau$.)
%
For this, we rewrite Eq.\ \eqref{eq:variational_radiusEq} as an equation for
$f(t)$.  For simplicity, we will write this out explicitly only in the limit
$U_{\rm i,f}\gg1$, and small oscillations, $f(t){\ll}R_0(t)$.  (The arguments
can of course be modified to go beyond the large-$U$ restriction. Small $f(t)$
is guaranteed for large $\tau$.)  We obtain
\begin{equation}
f''(t) + \Omega_B^2f(t) + \frac{u''(t)}{(D+2)u^{\frac{D+1}{D+2}}} -
\frac{(D+1)u'(t)^2}{(D+2)^2u^{\frac{2D+3}{D+2}}} = 0 ,
\end{equation}
with $u=U/(2\pi)^{D/2}$.  The first two terms give pure oscillations, i.e., breathing mode at fixed
$u$ with frequency $\Omega_B \approx \sqrt{D+2}\omega_{\mathrm{tr}}$.  The last two terms are
corrections due to time-varying interaction.
We first treat $\alpha>1$ ramps. The initial conditions at $t=0^+$ are then
$f(0)=f'(0)=0$.
With $u=u_{i}+ (\delta{u})(t/\tau)^{\alpha}$, the $u''$ correction is dominant
compared to the $u'^2$ correction at $t\ll\tau$.  The dominant correction
terms take the form $c_1/\tau^2$ for $\alpha=2$, and $c_1t/\tau^3$ for
$\alpha=3$.
%
%% Using $u=u_{i}+ (\delta{u})(t/\tau)^{\alpha}$, the last two terms above
%% take the forms $c_1/\tau^2-c_2t^2/\tau^4$ for $\alpha=2$, and
%% $c_1t/\tau^3-c_2t^4/\tau^6$ for $\alpha=3$.  
% 
The solutions of the resulting differential equation are sums of oscillatory
and algebraic terms.  It is straightforward to verify that the boundary
conditions $f(0)=f'(0)=0$ force the oscillatory part to have coefficients
scaling as $\sim\tau^{-\alpha}$.
% $\sim\tau^{-2}$ for $\alpha=2$ and $\sim\tau^{-3}$ for $\alpha=3$.  
%
This explains the $Q\sim\tau^{-2\alpha}$ behavior for integer $\alpha>1$.
The $\alpha=1$ case is slightly different.  The initial condition
still involves $\sigma'(0)=0$, but since $R_0(t)=[u(t)]^{1/(D+1)}$ has
finite slope at $t=0^+$, this now corresponds to
$f'(0^+)=-R_0'(0^+)=-c_3/\tau$. This initial condition leads to a
purely oscillatory $f(t)$ with amplitude $\sim\tau^{-1}$, which
explains $Q(\tau)\sim\tau^{-2}$ for $\alpha=1$.

\section{Spectral Interpretation \label{sec_spectral}}

For small changes of interaction, one can use the elegant perturbative results of
Ref.\ \cite{EcksteinKollar_NJP2010} to interpret our generic results in terms of the spectal
structure of many-body systems in harmonic traps.  The perturbative expression is
\begin{equation}
Q(\tau) \propto \int \frac{d\omega}{\omega} \Gamma(\omega) F(\omega\tau)
\end{equation}
where 
\begin{equation}
F(u)    ~=~ \left| \int_0^{x_{\rm max}} dx s'(x) e^{iux} \right|^2
\end{equation}
encodes the relevant information about the rampshape $s(x)$ ($x_{\rm max}$ is 1 for
[A] ramps and  $\infty$ for [B] ramps), and 
\begin{equation}
\Gamma(\omega) ~\propto~ \sum_{n\neq0} \left| \bra{\phi_n}\hat{W}
\ket{\phi_0} \right|^2 \delta(\omega-\epsilon_{n0})
\end{equation}
encodes the relevant spectral structure of the system \cite{EcksteinKollar_NJP2010}. 
Here $\ket{\phi_n}$'s are the eigenstates of the system, $\epsilon_{n0}$ are the eigenenergies
measured from the ground state energy, and $\hat{W}$ is the perturbing operator, i.e. the part of
the Hamiltonian being ramped, in our case the interaction term.
%
%% We have used the notation $\Gamma$ instead of the $R$ of
%% Ref.~\cite{EcksteinKollar_NJP2010} to avoid conflict.  

The eigenspectra of interacting systems in harmonic traps are not known in great detail, to the best
of our knowledge.  However, we can make the following general observations.  The lowest excited eigenstates
(at or around $\omega_{\rm tr}$) are spatially asymmetric, and are
therefore not excited by interaction ramps, due to symmetry.  (Excitation of these eigenstates would
lead to dipole mode oscillations of the cloud center of mass.)
The states at or around $2\omega_{\mathrm{tr}}$ are more relevant for interaction ramps.  Since
these spatially-symmetric eigenstates have radial size larger than the ground state, a small
component of these eigenstates in the wavefunction leads to breathing-mode oscillations.  Since the
interaction affects the ground-state radial size of the trapped system, at least one of these states
around $2\omega_{\mathrm{tr}}$ should be excited in an interaction ramp.  The excitation energy of
this eigenstate over the ground state is the breathing mode frequency $\Omega_B$.  

(Above, we conjectured the lowest antisymmetric and lowest symmetric excited states to lie around
$\omega_{\mathrm{tr}}$ and around $2\omega_{\mathrm{tr}}$ respectively, because that is the spectral
structure of the non-interacting gas.  With interactions, it may be reasonable to presume a similar
structure.  Our argument below does not depend on the exact values of the excitation energies.)

Since we are interested in large $\tau$ and are now considering the perturbative situation, we will
approximate the lowest spatially symmetric eigenstate to be the only excitation. i.e.,
$\Gamma(\omega) \propto \delta(\omega-\Omega_B)$.  With this approximation, the heat function is
simply
\begin{equation}
Q(\tau) \propto F(\Omega_B\tau) \ , 
\end{equation}
i.e., it is completely set by the ramp shape except for the breathing mode frequency $\Omega_B$
setting the scale for $\tau$.

\begin{figure}
\centering
\includegraphics*[width=0.99\columnwidth]{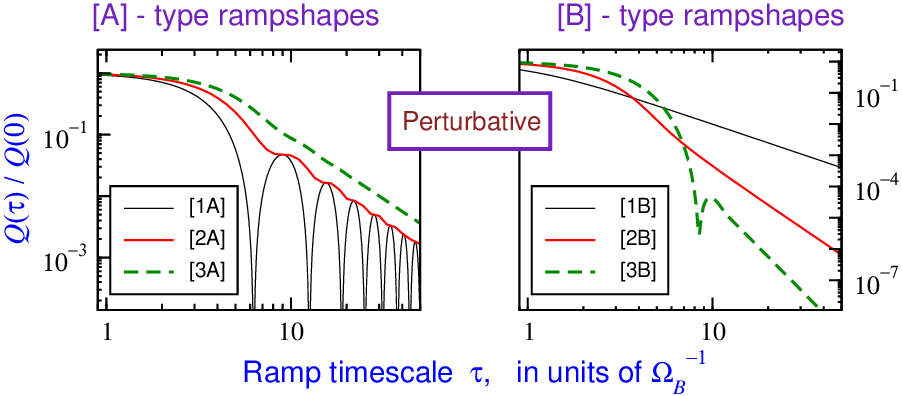}
\caption{ \label{fig_spectral}
Excess energy predictions from perturbative formalism \cite{EcksteinKollar_NJP2010} based on
spectral considerations (Sec.\ \ref{sec_spectral} in text).  
}
\end{figure}

Evaluating $F(u)$ for the different rampshapes, we find the following expressions for the normalized
excitation energy $Q(\tau)/Q(0)$:
\begin{align*}
&{\rm [1A]} \qquad \frac{4\sin^2(\ttau/2)}{\ttau^2} \\
&{\rm [2A]} \qquad \frac{4}{\ttau^2} -\frac{8\sin\ttau}{\ttau^3} + \frac{16\sin^2(\ttau/2)}{\ttau^4} \\ 
&{\rm [3A]} \qquad \frac{9}{\ttau^2} +\frac{36\cos\ttau}{\ttau^4} -\frac{72\sin\ttau}{\ttau^5} 
  + \frac{146\sin^2(\ttau/2)}{\ttau^6}   \\
&{\rm [1B]} \qquad  \frac{1}{1+\ttau^2}
\end{align*}
with $\ttau \equiv \Omega_B\tau$.  The [2B] and [3B] cases can be written in terms of special
functions; it is not helpful to write out these complicated expressions in full, but they have the
correct asymptotics, $\sim\ttau^{-4}$ ([2B]) and  $\sim\ttau^{-6}$ ([3B]).  Figure \ref{fig_spectral} plots these results.
Clearly, all the features presented in Sec.\ \ref{sec_universal_features} and Figure \ref{fig_heatcurves} are reproduced from this simplified
analysis.  The sole exeption is that the [1A] case would requre a non-oscillating $\tau^{-2}$ term
to match the forms found for the physical systems.  This can presumably be obtained by going beyond
perturbation theory.

\section{Discussion; Open questions \label{sec_discussion}}

\subsection{Summary and context}
 
Considering various harmonic-trapped atomic clouds, we have presented results on the adiabaticity
question, demonstrating system-independent aspects in the first corrections to adiabatic behavior
for slow ramps.
We have shown that the slow-ramp response is determined by the radius-oscillation modes common to
many trapped atomic systems, and its exact form depends on the ramp shape.  We have explained the
universal effects using a single-radius description of the cloud.  This covers a wide range of
interesting systems, but it is even more remarkable that our universal features extend to at least
some systems which cannot be described by a single radius.  In the perturbative regime (small
quenches), the generic features can be alternatively derived by using the formalism of
Ref.\ \cite{EcksteinKollar_NJP2010}, assuming a single excited eigenstate to determine the heating.
The connection between the two pictures is that the excited eigenstate is expected to be (one of)
the lowest spatially-symmetric eigenstates which has different size compared to the ground state;
hence excitation of this eigenstate leads to breathing-mode oscillations of the cloud size.

We have shown that a final kink in the ramp shape plays a drastic role in the non-adiabatic response
of compressible trapped systems.  A recently discovered effect of such kinks is logarithmic
contributions to $Q(\tau)$ \cite{Polovnikov_AdiabaticPertThy, EcksteinKollar_NJP2010,
  DoraHaqueZarand_arxiv10}.  The effect we have found for trapped systems (kink induces larger
oscillations overwhelming initial excitation) is quite different.

%% Oscillations of $Q(\tau)$ are relatively poorly understood, and may well be
%% generic in many-body ramps even without a trap.  We have provided a very
%% physical interpretation of $Q(\tau)$ oscillations in terms of radius
%% oscillations.  In other known examples of $Q(\tau)$
%% oscillations \cite{CanoviRossiniFazioSantoro_JSM09, EcksteinKollar_NJP2010,
%% Venumadhav_PRB2010}, the physical explanations, where known, are all
%% different.

\subsection{Open questions}

Our work opens up several new research avenues.  
Like other ``universal'' results, it is important to identify the limits of
validity.  For example, do the same $Q(\tau)$ asymptotic features appear in
trapped systems not described by a single radius?  Paradigm examples are
Bose-Hubbard systems containing superfluid-insulator ``wedding-cake''
structures, and phase-separated imbalanced Fermi gases near unitarity.  We
have shown some examples where the same behaviors appear
[Figs.\ \ref{fig_heatcurves}(c,f)], but a general understanding is lacking.
It is likely that the dynamics of one radius-like variable generally
dominates the extreme asymptotics, recovering our results.

Each of the systems are of intense interest in their own right, and understanding less universal
features in parameter quenches is important for the individual systems, especially with growing
experimental interest and capabilities for studying ramps and quenches.
Ramps in the trapping frequency should also induce radius oscillations, but details might be
different from interaction ramps.
The case of anisotropic harmonic traps also remains an open issue.  One might speculate that one of
the trapping frequencies dominate the extreme asymptotics, but the intermediate-$\tau$ region might
show interesting interplay of the several frequencies and associated radii.
Physical insights developed in our study of $Q(\tau)$ can perhaps be applied to better understand
``optimal ramp'' studies seeking to find ramp paths producing minimal heating
\cite{Muga_delCampo_frictionless}.
Finally, the spectral considerations of Sec.\ \ref{sec_spectral} highlight that the spectral
structure of many-body systems in the presence of an external harmonic trap deserves to be better
studied.

\appendix

\section{Calculation Methods and Approximations \label{sec_appendix}}

In this appendix we provide details on the methods and approximations used to obtain the $Q(\tau)$
results that are presented in Figure \ref{fig_heatcurves} for the different model systems.

\subsection{Lattice systems}

For the data of Fig.\ \ref{fig_heatcurves}(a-c) (fermionic and bosonic Hubbard models), we
time-evolved the full wavefunctions numerically exactly.  
The initial wavefunction was obtained by Lanczos diagonalization of the Hamiltonian with initial
interaction $U_{\rm i}$.  
For ramps (changing Hamiltonian), numerical time evolution of the full wavefunction is more
challenging than time-evolving under a constant Hamiltonian, for which efficient Krylov subspace
based methods exist \cite{evolution_Krylov}.
One option for ramps is to break the evolution into time-steps within which the Hamiltonian is
approximated to be constant.  Instead, we used Runge-Kutta evolution with adaptive stepsize.
Calculating $Q(\tau)$ at large $\tau$ (where $Q$ becomes small) requires high precision as $Q(\tau)$
is a difference between two energies that are close in value.  Because of these restrictions, the
Hilbert spaces for the data shown in Fig.\ \ref{fig_heatcurves}(a-c) were around ${\sim}10^5$.
Typically we used five to ten bosons or fermions in ten to fifteen sites.  Relatively strong traps,
$k_{\rm tr} \sim \ord(1)$, were used in order to ensure that the cloud edges did not reach the
lattice edges. 

For the Bose-Hubbard model, the exact evolution was complemented by calculations using the
Gutzwiller approximation, e.g., Fig.\ \ref{fig_heatcurves}(f), which allows for larger sizes.
This is a widely used approach for time evolution in Bose-Hubbard systems; see. e.g.,
Ref.\ \cite{Lewenstein_review} for a recent description.  The ansatz wavefunction is 
\begin{equation}
\ket{G(t)} ~=~ \prod_{i} \sum_{n} f_{n}^{(i)}(t) \ket{n;i}
\end{equation}
where $\ket{n;i}$ is the $n$-boson Fock state on site $i$.  The $f_{n}^{(i)}$ parameters for a
particular site $i$ may be regarded as the components of a local wavefunction, which evolve
according to the local site Hamiltonian
\begin{multline}
\hat{H}_i  ~=~ (V_i-\mu) n_i + {\hf}Un_i(n_i-1) \\ - {b}_i\sum_{<j>}\Phi_j^* 
- {b}_i^{\dagger}\sum_{<j>}\Phi_j  ,  
\end{multline}
where $V_i= {\hf}k_{\rm tr}i^2$ is the trap potential, the $j$ index runs over all sites neighboring
the $i$ site, and
\[
\Phi_j ~=~ \xpct{b_j} ~=~ \sum_{n} \sqrt{n+1}\,  f_{n}^{(j)*}  f_{n+1}^{(j)}
\]
is the condensate fraction at site $j$.  As in the case of full wavefunction evolution, Gutzwiller
evolution is much more demanding for a changing Hamiltonian compared to evolution under a constant
Hamiltonian, even more so in our case because of the high precision required for the final energy in
order to calculate $Q(\tau)$.  We employed imaginary time evolution to find the initial ground
state, and then Runge-Kutta (with adaptive timestep size) to evolve the coupled equations for
$f_{n}^{{i}}(t)$ in real time with changing $U(t)$.

\subsection{Continuum systems}

\paragraph*{Bose condensate.}

In addition to numerical solutions of the GP equation, we have also used a single-parameter
variational description to describe the dynamics, using the condensate size $\sigma(t)$ as the
time-dependent parameter.  This description is particularly convenient for analyzing breathing-mode
dynamics, as we have done in Section \ref{sec_radius_dynamics_analysis}.  

The radius description is formulated in terms of a Gaussian variational ansatz, which for 1D is
\begin{equation}  \label{eq:Ansatz1}
\psi(x,t)=\frac{1}{[\sqrt{\pi}\sigma(t)]^{1/2}}\exp
\left[
-\frac{x^2}{2[\sigma(t)]^2}-i\beta(t) x^2
\right] .
\end{equation}
For $D>1$ the variational wave function is a product of one such Gaussian factor for each dimension.
Using this ansatz in the GP Lagrangian
\begin{equation}   \label{eq:GP_lagrangian}
L= \frac{i}{2} \int_r  \left(
\psi^*\frac{\partial\psi}{\partial{t}} -\psi\frac{\partial\psi^*}{\partial{t}}  
%\psi^*\partial_{t}\psi -\psi\partial_{t}\psi^*
\right) -E[\psi] \  ,
\end{equation}
the Euler-Lagrange equations of motion give the evolution
equations for the variational parameters $\sigma(t)$ and $\beta(t)$. 
%
%% \cite{PerezGarcia_Cirac_Lewenstein_Zoller_variational}.
%
This is a standard and widely-used technique for GP dynamics, dating back to
Ref.~\cite{PerezGarcia_Cirac_Lewenstein_Zoller_variational}.  

The imaginary part in the wave function \eqref{eq:Ansatz1} is necessary because time evolution from
a real wave function produces an imaginary component.  However, the two parameters turn out to be
not independent but simply related ($\beta(t) \propto\del_t\ln\sigma(t)$).  There is thus
effectively a single dynamical parameter describing the system, namely the cloud radius $\sigma(t)$.
The resulting equation of motion for $\sigma$ is found to be Eq.\ \eqref{eq:variational_radiusEq},
and the energy in terms of $\sigma$ is given by Eq.\ \eqref{eq:variational_energyEq}.

In comparing the single-parameter variational description with the full GP equations,, we find that
the normaized excess energy, $Q(\tau)/Q(0)$, is reproduced almost exactly by the single-parameter
description, but that the $Q(\tau)$ values obtained from  Eqs.\ \eqref{eq:variational_radiusEq},
\eqref{eq:variational_energyEq}  do not have the correct normalization and deviate by a factor from
full-GP results.  Note that the normalization does not affect any of the universal behaviors
(Sec.\ \ref{sec_universal_features}, Fig.\ \ref{fig_heatcurves}) under discussion in this Article.

We could just as well use a Thomas-Fermi instead of Gaussian profile.  The results are very similar
and do not substantially affect any of the arguments we make in this work.  The equation of motion
in that case loses the $\sigma^{-2}$ term of Eq.\ \eqref{eq:variational_radiusEq}
(Sec.\ \ref{sec_systems}).

\paragraph*{Continuum two-component Fermi system.}

To perform non-equilibrium calculations for two-component (spin-$\hf$) continuum fermionic gases, we
have used a quantum hydrodynamic approximation.  The formulation of Refs.\ \cite{fermions_tddft} in
terms of a nonlinear Schr\"odinger equation (``time-dependent DFT'') is convenient for our purposes
because of its slimiarity to the GP description of continuum bosons.

The time-dependent DFT formulation has been used successfully for the unitary Fermi gas (reviewed in
Ref.\ \cite{Bulgac_1301_review}).  We do not consider the unitary limit in this work because the
interaction is fixed at infinity and thus cannot be ramped as a time-dependent parameter.  We focus
on weakly interacting gases so that we can use the relatively simple Hartree approximation and the
interaction $U(t)$ appears explicitly and can be ramped.  In this work we have restricted ourselves
to 3D and to equal populations of the two components.

Using the Hartree approximation for the chemical potential (main text
Eq.\ \eqref{eq:Hartree_chempot}), the hydrodynamic equation becomes
\begin{equation}
i\hbar \partial_t\psi
=
\left[-\frac{\hbar^2}{2 m}\nabla^2\psi+V_{\rm ext}+\frac{U(t)}{2}|\psi|^2+\alpha|\psi|^{4/3} \right]\psi  
\end{equation}
with $ \alpha = \hf \lba 3\pi^2\rba^{2/3}$, and $n=|\psi|^2$ here is the total density of the two
spin states.   Note the interaction being ${\hf}Un$ rather than $Un$.  Each fermion interacts
with half of all the fermions, those with the opposite spin. 

In contrast to the Bose condensate case, it is not consistent to normalize $\psi$ to unity.  This
means that there are no factors of $N$ absorbed in the definition of $U$ in the continuum fermionic
system, as opposed to the continuum Bose condensate.  This difference is responsible for the rather
different ranges of values for the $U(t)$ of BECs in Figure \ref{fig_heatcurves}(d) and the $U(t)$
of the 3D Fermi gas in Figure \ref{fig_heatcurves}(e).

As in the Bose condensate case, we employ the variational ansatz
\begin{align}
\psi({\bf r},t) ~=~ \frac{\sqrt{N}}{\sqrt{\pi^{3/2}\sigma^3}} 
\exp\left[\frac{1}{2}\left(\frac{r}{\sigma}\right)^2-i\beta r^2\right]. 
\end{align}
Note the factor $\sqrt{N}$.  The contrast to Eq.\ \eqref{eq:Ansatz1} is due to the different
normalization.  The Euler-Lagrange equations provide equations of motion for $\sigma(t)$ and
$\beta(t)$.  The two parameters are not independent ($\beta=-\frac{1}{2}\frac{\dot\sigma}{\sigma}$,
identical to the GP case).  Eliminating $\beta$, we get a single-parameter description.  The
equation of motion for $\sigma$ is
\begin{equation}
\sigma\ddot\sigma+\sigma^2-\frac{1}{\sigma^2}-\frac{UN}{2(\sqrt{2\pi}\sigma)^3}
-\tfrac{4}{3}\left(\tfrac{3}{5}\right)^{5/2}\frac{N^{2/3}\alpha}{\pi\sigma^2}
= 0 \ ,
\end{equation}
and the energy equation is 
\begin{equation}
E=\frac{3N}{4}\left[\frac{1}{\sigma^2}+\dot\sigma^2+\sigma^2\right]
+\frac{1}{(2\pi)^{3/2}}\frac{UN^2}{4\sigma^3}
+\left(\tfrac{3}{5}\right)^{5/2}\frac{N^{5/3}\alpha}{\pi\sigma^2}.
\end{equation}

\begin{acknowledgments}

MH thanks M.~Snoek for discussion on implementing the Gutzwiller
approximation.

\end{acknowledgments}


\begin{thebibliography}{99}




\bibitem{finiteRateQuenches_reviews}
%
See review and citations of earlier work in:  
%
\\ J.~Dziarmaga, Adv.\ Phys.\ {\bf 59}, 1063 (2010); 
%
\\ A.~Polkovnikov, K.~Sengupta, A.~Silva, M.~Vengalattore,  Rev.\ Mod.\ Phys.\ {\bf 83}, 863 (2011).  
%
\\ Chapters 2-5 in \emph{Quantum Quenching, Annealing and Computation}, edited
by A.~K.~Chandra, A.~Das, and B.~K.~Chakrabarti, Springer, 2010.

	
\bibitem{Polovnikov_AdiabaticPertThy}
C.~De~Grandi, V.~Gritsev, and A.~Polkovnikov, Phys.\ Rev.\ B {\bf 81}, 012303
(2010); {\bf 81}, 224301 (2010).
%


%% \bibitem{finiteRateQuenches_othersystems} 
%% %
%% For recent work, see, \emph{e.g.}, 
%% %
%% A.~Rahmani and C.~Chamon, arXiv:1011.3061.
%% %
%% N.~Eurich, M.~Eckstein, and P.~Werner, 	arXiv:1010.2853.  
%% %
%% C.-C.~Chien and B.~Damski, Phys.\ Rev.\ A {\bf 82}, 063616 (2010).
%% %
%% C.~De~Grandi, V.~Gritsev, and A.~Polkovnikov, Phys.\ Rev.\ B {\bf 81}, 012303 (2010);  {\bf 81}, 224301 (2010).
%% %
%% F.~Pollmann, S.~Mukerjee, A.~G.~Green, and J.~E.~Moore, Phys. Rev. E 81, 020101 (2010).
%% %
%% K.~Sengupta and D.~Sen, Phys.\ Rev.\ A {\bf 80}, 032304 (2009).
%% %
%% S.~Miyashita, H.~De Raedt, and B.~Barbara, Phys.\ Rev.\ B {\bf 79}, 104422 (2009).  
%% %
%% U.~Divakaran and A.~Dutta, 
%% Phys.\ Rev.\ B {\bf 79}, 224408 (2009). 
%% %
%% S.~Mondal, K.~Sengupta, and D.~Sen, Phys.\ Rev.\ B {\bf 79}, 045128 (2009).
%% %
%% A.~Bermudez, D.~Patan\`e, L.~Amico, and M.~A.~Martin-Delgado,
%% Phys.\ Rev.\ Lett.\ {\bf 102}, 135702 (2009).
%% %
%% L.~Cincio, J.~Dziarmaga, J.~Meisner, and M.~M.~Rams, 
%% Phys.\ Rev.\ B {\bf 79}, 094421 (2009).
%% %
%% D.~Patan\`e {\it et.~al.},
%% % D.~Patan\`e, L.~Amico, A.~Silva, R.~Fazio, and G.~E.~Santoro, 
%% Phys.\ Rev.\ Lett.\ {\bf 101}, 175701 (2008).  
%% %
%% S.~Mondal, D.~Sen, and K.~Sengupta, Phys.\ Rev.\ B {\bf 78}, 045101 (2008). 
%% %
%% R.~Sch\"utzhold, J.~Low Temp.\ Phys.\ {\bf 153}, 228 (2008).
%% %
%% R.~Barankov and A.~Polkovnikov, Phys.\ Rev.\ Lett.\ {\bf 101},076801 (2008).
%% %
%% D.~Sen, K.~Sengupta, and S.~Mondal, Phys.\ Rev.\ Lett.\ {\bf 101}, 016806 (2008). 
%% %
%% K.~Sengupta, D.~Sen, and S.~Mondal, Phys.\ Rev.\ Lett.\ {\bf 100}, 077204 (2008).
%% %
%% V.~Mukherjee, U.~Divakaran, A.~Dutta, and D.~Sen, Phys.\ Rev.\ B {\bf 76},
%% 174303 (2007). 
%% %
%% M.~Uhlmann, R.~Sch\"utzhold, and U.~R.~Fischer, Phys.\ Rev.\ Lett.\ {\bf 99},
%% 120407 (2007). 
%% %
%% A.~Das, K.~Sengupta, D.~Sen, and B.~K.~Chakrabarti, Phys.\ Rev.\ B {\bf 74},
%% 144423 (2006).
%% %
%% R.~Sch\"utzhold, M.~Uhlmann, Y.~Xu, and U.~R.~Fischer, Phys.\ Rev.\ Lett.\ {\bf 97},
%% 200601 (2006).
%% %
%% W.~H.~Zurek, U.~Dorner, and P.~Zoller, Phys.\ Rev.\ Lett.\ {\bf 95}, 105701 (2005).
%% %
%% J.~Dziarmaga, Phys.\ Rev.\ Lett.\ {\bf 95}, 245701 (2005).



\bibitem{finiteRateQuenches_residualenergy}
%
T.~Caneva, R.~Fazio, and G.~E.~Santoro, Phys.\ Rev.\ B {\bf 76}, 144427
(2007).
%
F.~Pellegrini, S.~Montangero, G.~E.~Santoro, and R.~Fazio, Phys.\ Rev.\ B {\bf
77}, 140404(R) (2008).
%
T.~Caneva, R.~Fazio, and G.~E.~Santoro, Phys.\ Rev.\ B {\bf 78}, 104426 (2008).


%% \bibitem{CucchiettiDamskiDziarmagaZurek_BoseHubbardDynamics_PRA07}
%% F.~M.~Cucchietti, B.~Damski, J.~Dziarmaga, and W.~H.~Zurek,
%% Phys.\ Rev.\ A {\bf 75}, 023603 (2007).
%% %
%% % \\ \emph{Dynamics of the Bose-Hubbard model: Transition from a Mott insulator to a superfluid.}


\bibitem{EcksteinKollar_NJP2010} M.~Eckstein and M.~Kollar, New J.~Phys.\ {\bf
12}, 055012 (2010).



\bibitem{CanoviRossiniFazioSantoro_JSM09} E.~Canovi, D.~Rossini, R.~Fazio, and
  G.~E.~Santoro, J.~Stat.\ Mech.\ (2009) P03038.

\bibitem{MoeckelKehrein_NJP2010}
M.~Moeckel and S.~Kehrein,  New J.~Phys.\ {\bf 12}, 055016 (2010). 

\bibitem{DoraHaqueZarand_arxiv10}
B.~ D\'ora, M.~Haque, and G.~Zar\'and,   Phys.\ Rev.\ Lett.\ {\bf 106}, 156406 (2011). 

\bibitem{DziarmagaTylutki_arxiv11}
J.~Dziarmaga and M.~Tylutki,  arXiv:1109.3801.  
%
% \\ \emph{Excitation energy after a smooth quench in a Luttinger liquid}
    

\bibitem{Venumadhav_PRB2010} T.~Venumadhav, M.~Haque, and R.~Moessner,\, Phys.\ Rev.\ B {\bf 81}, 054305 (2010).  


\bibitem{ramp_papers_finitesystems}  
%
G.~Roux, Phys.\ Rev.\ A {\bf 81}, 053604 (2010).
%
J.-S.~Bernier, G.~Roux, and C.~Kollath,  Phys.\ Rev.\ Lett.\ {\bf 106}, 200601 (2011).
%
M.~Collura and D.~Karevski,  Phys.\ Rev.\ Lett.\ {\bf 104}, 200601 (2010). 
%
S.S.~~Natu, K.~R.~A.~Hazzard, and E.~J.~Mueller, Phys.\ Rev.\ Lett.\ {\bf 106}, 125301 (2011).

%% \bibitem{Roux_PRA2010} G.~Roux, Phys.\ Rev.\ A {\bf 81}, 053604 (2010).

%% \bibitem{BernierRouxKollath_PRL11} 
%% J.-S.~Bernier, G.~Roux, and C.~Kollath,  Phys.\ Rev.\ Lett.\ {\bf 106}, 200601 (2011).
%% % arXiv:1010.5251.  
%% %
%% % \\ \emph{Slow quench dynamics of a trapped one-dimensional Bose gas confined to an optical lattice}

%% \bibitem{ColluraKarevski_PRL10} M.~Collura and D.~Karevski,  
%% Phys.\ Rev.\ Lett.\ {\bf 104}, 200601 (2010). 
%% %
%% % \\ \emph{Critical Quench Dynamics in Confined Systems}


%% \bibitem{CampostriniVicari_arxiv10} 
%% M.~Campostrini and E.~Vicari, arXiv:1010.0806. 
%
% \\ \emph{Equilibrium and off-equilibrium trap-size scaling in 1D ultracold bosonic gases}



% \bibitem{AdiabaticQC} E.~Farhi, J.~Goldstone,  S.~Gutmann, and  M.~Sipser, Science {\bf 292}, 472 (2001); arXiv:quant-ph/0001106.



\bibitem{Salomon_BEC_ramp_2011} 
%
% N.~Navon \textit{et al},   arxiv:1103.4449.
%
N.~Navon, S.~Piatecki, K.~J.~G\"unter, B.~Rem, T.~C.~Nguyen, F.~Chevy,
W.~Krauth, and C.~Salomon,  arxiv:1103.4449.
%
%\\ \emph{Dynamics and Thermodynamics of the Low-Temperature Strongly Interacting Bose Gas.}    



\bibitem{HungZhangGemelkeChin_PRL10} C.~L.~~Hung, X.~Zhang, N.~Gemelke, and
  C.~Chin,  Phys.\ Rev.\ Lett.\ {\bf 104}, 160403 (2010).


\bibitem{BakrPolletGreiner_Science2010}  W.~S.~Bakr \textit{et al}., 
%
% W.~S.~Bakr, A.~Peng, M.~E.~Tai, R.~Ma, J.~Simon,  J.~I.~Gillen, S.~F\"olling,
% L.~Pollet, and M.~Greiner, 
%
Science {\bf 329}, 547 (2010).


\bibitem{DeMarco_PRL11}
D.~Chen, M.~White, C.~Borries, and B.~DeMarco, Phys.\ Rev.\ Lett.\ {\bf 106}, 235304 (2011). 
%
% \\ \emph{Quantum Quench of an Atomic Mott Insulator.}


\bibitem{Lewenstein_review} 
%
M.~Lewenstein, A.~Sanpera, V.~Ahufinger, B.~Damski, A.~Sen(De), and U.~Sen,
% M.~Lewenstein \textit{et.~al.},
Adv.\ Phys.\ {\bf 56}, 243 (2007).
%
Hubbard models in the cold-atom context are described in Section 2.  The
Gutzwiller approximation is reviewed in Section 3.4.



%% \bibitem{gutz} 
%% %
%% The Gutzwiller approximation is reviewed in Section 3.4 of:
%% % M.~Lewenstein, A.~Sanpera, V.~Ahufinger, B.~Damski, A.~Sen (De), and U.~Sen,
%% M.~Lewenstein \textit{et.~al.},
%% Adv.\ Phys.\ {\bf 56}, 243 (2007).
%% 
%% D.~S.~Rokhsar and B.~G.~Kotliar, Phys.\ Rev.\ B {\bf 44}, 10328 (1991).
%% %
%% W.~Krauth, M.~Caffarel, and J.-P.~Bouchaud, Phys.\ Rev.\ B {\bf 45}, 3137 (1992).
%% %
%% K.~Sheshadri, H.~R.~Krishnamurthy, R.~Pandit, and T.~V.~Ramakrishnan,
%% Europhys.\ Lett.\ {\bf 22}, 257 (1993).
%% %
%% D.~Jaksch, C.~Bruder, J.~I.~Cirac, C.~W.~Gardiner, and P.~Zoller,
%% Phys.\ Rev.\ Lett.\ {\bf 81}, 3108 (1998).
%% %
%% D.~Jaksch, V.~Venturi, J.~I.~Cirac, C.~J.~Williams, and P.~Zoller, Phys.\
%% Rev.\ Lett.\ {\bf 89}, 040402 (2002).
%% %
%% B.~Damski, J.~Zakrzewski, L.~Santos, P.~Zoller, and M.~Lewenstein, Phys.\
%% Rev.\ Lett.\ {\bf 91}, 080403 (2003). 
%% %
%% J.~Zakrzewski, Phys.\ Rev.\ A {\bf 71}, 043601 (2005).
%% %
%% X.~Lu and Y.~Yu, Phys.\ Rev.\ A {\bf 74}, 063615 (2006).
%% %
%% M.~Snoek and W.~Hofstetter, Phys.\ Rev.\ A {\bf 76}, 051603(R) (2007).
%% %
%% C.~Trefzger, C.~Menotti, and M.~Lewenstein, Phys. Rev. A 78, 043604 (2008). 
%% %
%% P.~Navez and R.~Schützhold, Phys.\ Rev.\ A {\bf 82}, 063603 (2010).


%% \bibitem{KibbleZurek} T.~W.~B.~Kibble, J.~Phys.\ A {\bf 9}, 1387 (1976);
%% Phys.\ Rep.\ {\bf 67}, 183 (1980). \;
%% %
%% W.~H.~Zurek, Nature  {\bf 317}, 505 (1985); Phys.\ Rep.\ {\bf 276}, 177
%% (1996).


\bibitem{Muga_delCampo_frictionless} J. G. Muga, Xi Chen, A. Ruschhaupt, and
  D. Guery-Odelin, J.~Phys.~B {\bf 42}, 241001 (2009). 
%
% \\ \emph{Frictionless dynamics of Bose-Einstein condensates under fast trap variations}
%
A.~del~Campo, Phys.\ Rev.\ A {\bf 84}, 031606(R) (2011). 
%
% \\ \emph{Frictionless quantum quenches in ultracold gases: a quantum dynamical microscope.}


\bibitem{pitaevskii-jetp13} L. P. Pitaevskii, Sov. Phys. JETP {\bf 13}, 451
  (1961). 
% Zh. Eksp. Teor. Fiz. {\bf 40}, 646 (1961)[Sov. Phys. JETP {\bf 13}, 451 (1961)].

\bibitem{gross-nc20}  E. P. Gross, Nuovo Cimento {\bf 20}, 454 (1961). 


\bibitem{PerezGarcia_Cirac_Lewenstein_Zoller_variational}
V.~M.~Perez-Garcia, H.~Michinel, J.~I.~Cirac, M.~Lewenstein and P.~Zoller,
% V.~M.~Perez-Garcia \textit{et.\ al.}, 
Phys.\ Rev.\ Lett.\ {\bf 77}, 5320 (1996);  Phys.\ Rev.\ A {\bf 56}, 1424 (1997).


\bibitem{Olshanii_PRL98}  M.~Olshanii, Phys.\ Rev.\ Lett.\ {\bf 81}, 938 (1998).



% \bibitem{MenottiPedriStringari_PRL02} C.~Menotti, P.~Pedri, and
% S.~Stringari, Phys.\ Rev.\ Lett.\ {\bf 89}, 250402 (2002).

\bibitem{fermions_tddft} Y.~E.~Kim and A.~L.~Zubarev, Phys.\ Rev.\ A {\bf 70}, 033612 (2004). 
%
S.~K.~Adhikari,  Phys.\ Rev.\ A {\bf 77}, 045602 (2008). 
% 	
S.~K.~Adhikari,  J.~Phys.\ B {\bf 43}, 085304 (2010).  
%
P.~Díaz, D.~Laroze, I.~Schmidt, and B.~A.~Malomed,  J.~Phys.\ B {\bf 45}, 145304 (2012). 

\bibitem{evolution_Krylov}
T.~J.~Park and J.~C.~Light, J.~Chem.\ Phys.\ {\bf 85}, 5870 (1986). \\ 
S.~R.~Manmana, A.~Muramatsu, and R.~M.~Noack, AIP Conf.\ Proc.\ {\bf 789}, 269 (2005).

\bibitem{Bulgac_1301_review}
A.~Bulgac,  arXiv:1301.0357. 


\end{thebibliography}
\end{document}